\documentclass[letter]{aa}

\usepackage{graphicx}
\usepackage{txfonts}

\newcommand{\ha}{H$\alpha$}
\newcommand{\nii}{[\ion{N}{ii}]$\lambda$}
\newcommand{\hb}{H$\beta$}
\newcommand{\oiii}{[\ion{O}{iii}]$\lambda$}
\newcommand{\hg}{H$\gamma$}

\begin{document}

\title{Ratio of black hole to galaxy mass of an extremely red dust-obscured galaxy at $z = 2.52$}

\author{K. Matsuoka\inst{1,2,3}
        \and
        Y. Toba\inst{3,4,5}
        \and
        M. Shidatsu\inst{6}
        \and
        Y. Ueda\inst{5}
        \and
        K. Iwasawa\inst{7}
        \and
        Y. Terashima\inst{6}
        \and
        M. Imanishi\inst{8}
        \and
        T. Nagao\inst{3}
        \and
        A. Marconi\inst{1,2}
        \and
        W.-H. Wang\inst{4}
       }

\institute{Dipartimento di Fisica e Astronomia, Universit{\'a} degli Studi di Firenze,
           Via G. Sansone 1, I-50019 Sesto Fiorentino, Italy\\
           \email{matsuoka@arcetri.astro.it}
           \and
           INAF -- Osservatorio Astrofisico di Arcetri,
           Largo Enrico Fermi 5, I-50125 Firenze, Italy
           \and
           Research Center for Space and Cosmic Evolution,
           Ehime University, 2-5 Bunkyo-cho, Matsuyama 790-8577, Japan
           \and
           Academia Sinica Institute of Astronomy and Astrophysics,
           No. 1, Section 4, Roosevelt Rord, Taipei 10617, Taiwan
           \and
           Department of Astronomy, Kyoto University,
           Kitashirakawa-Oiwake-cho, Sakyo-ku, Kyoto 606-8502, Japan
           \and
           Graduate School of Science and Engineering, Ehime University,
           2-5 Bunkyo-cho, Matsuyama 790-8577, Japan
           \and
           ICREA and Institut de Ci{\`e}ncies del Cosmos, Universitat de Barcelona,
           Mart{\'i} i Franqu{\`e}s, 1, 08028 Barcelona, Spain
           \and
           National Astronomical Observatory of Japan (NAOJ),
           National Institutes of Natural Sciences (NINS), 2-21-1 Osawa, Mitaka, Tokyo 181-8588, Japan
           }

\abstract{
We present a near-infrared (NIR) spectrum of WISE J104222.11$+$164115.3, an extremely red dust-obscured galaxy (DOG), which has been observed with the Long-slit Intermediate Resolution Infrared Spectrograph (LIRIS) on the 4.2m William Hershel Telescope.
This object was selected as a hyper-luminous DOG candidate at $z \sim 2$ by combining the optical and IR photometric data based on the Sloan Digital Sky Survey (SDSS) and Wide-field Infrared Survey Explorer (WISE), although its redshift had not yet been confirmed.
Based on the LIRIS observation, we confirmed its redshift of 2.521 and total IR luminosity of $\log (L_{\rm IR}/L_\odot) = 14.57$, which satisfies the criterion for an extremely luminous IR galaxy (ELIRG).
Moreover, we indicate that this object seems to have an extremely massive black hole with $M_{\rm BH} = 10^{10.92} M_\odot$ based on the broad \ha\ line: the host stellar mass is derived as $M_\star$ = 10$^{13.55} M_\odot$ by a fit of the spectral energy distribution.
Very recently, it has been reported that this object is an anomalous gravitationally lensed quasar based on near-IR high-resolution imaging data obtained with the {\it Hubble Space Telescope.}  Its magnification factor has also been estimated with some uncertainty (i.e., $\mu =$ 53--122).
We investigate the ratio of the black hole to galaxy mass, which is less strongly affected by a lensing magnification factor, instead of the absolute values of the luminosities and masses.
We find that the $M_{\rm BH}/M_\star$ ratio (i.e., 0.0140--0.0204) is significantly higher than the local relation, following a sequence of unobscured quasars instead of obscured objects (e.g., submillimeter galaxies) at the same redshift.
Moreover, the LIRIS spectrum shows strongly blueshifted oxygen lines with an outflowing velocity of $\sim 1100$ km/s, and our Swift X-ray observation also supports that this source is an absorbed AGN with an intrinsic column density of $N_{\rm H}^{\rm int} = 4.9 \times 10^{23}$ cm$^{-2}$.
These results imply that WISE J104222.11$+$164115.3 is in a blow-out phase at the end of the buried rapid black hole growth.
}


\maketitle

\section{Introduction}
The coevolution of galaxies and supermassive black holes (SMBHs) has in recent decades received great attention since the tight positive correlation between the masses of black holes at the center of galaxies ($M_{\rm BH}$) and their hosting stellar spheroids ($M_{\rm bul}$) was found in the local Universe \citep[e.g.,][]{1995ARA&A..33..581K,1998AJ....115.2285M,2003ApJ...589L..21M,2009ApJ...698..198G,2013ApJ...772...49W}.
It is widely accepted that the formation and evolution of SMBHs and their host galaxies are related.
However, we do not have any clear picture of how they have coevolved in the cosmic history. In order to understand the mechanism of the coevolution, the $M_{\rm BH}$-$M_{\rm bul}$ relation at high redshift has been investigated in addition to the local relation, most often using galaxies with active galactic nuclei, AGNs \citep[e.g.,][]{2003ApJ...583..124S,2006MNRAS.368.1395M,2006ApJ...649..616P,2010ApJ...708..137M,2013ApJ...767...13S,2018ApJ...854...97D}.

By comparing the $K$-band magnitude and virial black hole mass of radio-loud broad-line AGNs at $0 < z < 2$, \citet{2006MNRAS.368.1395M} found that the ratio of black hole to galaxy mass evolves with redshift as $M_{\rm BH}/M_\star \propto (1+z)^{2.07}$.
This is consistent with some theoretical predictions of $M_{\rm BH}/M_\star \propto (1+z)^{1.5-2.5}$ \citep[e.g.,][]{2003ApJ...595..614W,2005ApJ...627L...1A}.
Based on a sample of 31 gravitationally lensed and 20 non-lensed AGNs at $1 < z < 4.5$, \citet{2006ApJ...649..616P} found that the $M_{\rm BH}/M_\star$ ratio becomes 3--6 times higher at $z \sim 2$ than the current ratio \citep[see also][]{2006ApJ...640..114P}, consistent with the evolving $M_{\rm BH}/M_\star$ ratio in \citet{2006MNRAS.368.1395M}.
Moreover, \citet{2012MNRAS.420.3621T} confirmed that this evolving $M_{\rm BH}/M_\star$ ratio is followed even at $z > 4$ with deep $K$-band imaging of the most luminous $z \sim 4$ quasars and the $z = 6.41$ quasar.
These results indicate that the $M_{\rm BH}/M_\star$ ratio of AGNs would evolve positively with redshift, although dispersions are large \citep[see also][]{2008ApJ...681..925W,2010ApJ...708..137M,2015Sci...349..168T}.

We note, however, that the redshift evolution of the $M_{\rm BH}/M_\star$ ratio described above was investigated using AGNs in the relatively evolved unobscured phase, instead of using obscured objects in a rapid evolutionary phase.
\citet{1988ApJ...325...74S} suggested a major-merger evolutionary scenario in which galaxy mergers induce a rapid starburst and an obscured black hole growth, followed by an unobscured phase after the gas is blown out \citep[see also][]{2012NewAR..56...93A}. To understand the $M_{\rm BH}/M_\star$ evolution correctly, it is crucial to focus on heavily obscured AGNs, for instance, submillimeter galaxies (SMGs) and ultra-luminous infrared (IR) galaxies (ULIRGs), which would show violent star formation (SF) with buried black hole growth.
\citet{2008AJ....135.1968A} reported that the $M_{\rm BH}/M_\star$ ratios of SMGs at $2.0 < z < 2.6$ are a factor of $\sim 3$ times lower than those found in comparably massive normal galaxies in the local Universe: $\sim 10$ times lowr than those predicted for luminous quasars and radio AGNs at $z \sim 2$ in \citet{2006ApJ...649..616P} and \citet{2006MNRAS.368.1395M}.
Moreover, they found that local ULIRGs are also located below the local $M_{\rm BH}/M_\star$ ratio.
Using a sample of X-ray obscured, dust-reddened quasars at $1.5 < z < 2.6$, \citet{2014MNRAS.443.2077B} claimed that the $M_{\rm BH}/M_\star$ ratios of obscured red quasars seem to increase with redshifts lower than what has been found for blue quasars.
Recently, by using 25 millimeter galaxies at $z =$ 1.5--3, \citet{2018ApJ...853...24U} suggested that the black hole masses of a large portion of star-forming galaxies hosting AGNs are lower than those expected from the local $M_{\rm BH}/M_\star$ ratio.
These results indicate that obscured populations may represent a rapid black hole growing phase immediately precede the blue quasars that are typically selected in optical surveys.
However, since black hole mass estimates of obscured objects are usually difficult, the evolution of $M_{\rm BH}/M_\star$ in obscured AGNs is still poorly understood.

In this letter, we present a near-IR spectrum of an extremely red dust-obscured galaxy (DOG), WISE J104222.11$+$164115.3 (hereafter referred to as WISE J1042$+$1641), selected with a new efficient method \citep{2016ApJ...820...46T}, and we investigate the ratio of its black hole to  the galaxy mass.
In Sect.~2 we describe our target selection, the IR and X-ray observations, and show the results.
We discuss the interpretation and the $M_{\rm BH}/M_\star$ evolution in Sect.~3.
Throughout this work, we assume $H_0 = 70$ km s$^{-1}$ Mpc$^{-1}$, $\Omega_\Lambda = 0.7$, and $\Omega_{\rm M} = 0.3$.

\section{Observations and results}

\subsection{Target selection}

Our target WISE J1042$+$1641 (R.A. = 160.5921545\degr, Dec. = $+$16.6875855\degr, J2000.0) was selected as a candidate of a hyper-luminous IR galaxy (HyLIRG) at $z \sim 2$ with a new effective method using the optical and IR catalogs obtained from the Sloan Digital Sky Survey \citep[SDSS;][]{2000AJ....120.1579Y} and the Wide-field Infrared Survey Explorer \citep[WISE;][]{2010AJ....140.1868W}, reported in \citet{2016ApJ...820...46T}.
By combining the SDSS Data Release 12 spectroscopic catalog and the AllWISE catalog, they performed a search for hyper-luminous DOGs by adopting the SDSS $i'$ band and WISE 22 $\mu$m color, $i' - [22],$ in AB magnitude.
Here we define DOGs as galaxies with $i' - [22] > 7.0$ \citep[see][]{2015PASJ...67...86T}.
The authors found that the $i' - [22]$ color correlates with the total IR luminosity and a high portion ($\sim$ 73\%) of galaxies with $i' - [22] > 7.4$ has $\log(L_{\rm IR}/L_\odot) > 13$ \citep[see also][]{2018ApJ...857...31T}.
Using this method, we selected about 1800 HyLIRG candidates (including WISE J1042$+$1641).
In particular, WISE J1042$+$1641 has $i' - [22] = 7.93$, indicating that the bulk of the optical and UV emission from AGN and/or SF is absorbed by the very dense surrounding dust.
Therefore, WISE J1042$+$1641 with its extremely red color is a good candidate of a heavily obscured HyLIRG.

\subsection{Near-infrared observation with LIRIS/WHT}

We obtained the near-infrared spectrum of WISE J1042$+$1641 using the Long-slit Intermediate Resolution Infrared Spectrograph \citep[LIRIS;][]{1998SPIE.3354..448M} on the 4.2m William Herschel Telescope (17A-C41: 14--15 February 2017) to determine its redshift.
The observation was performed with the {\tt lr\_hk} grism ($R = 700$) and the {\tt hkspec} filter to cover a wavelength range of 14000 \AA\ $< \lambda_{\rm obs} < 25000$ \AA, into which H$\alpha$ and H$\beta$ lines are redshifted if $z \sim 2$.
When we used the 1\farcs0 slit width, the final spectral resolution was 27.5 \AA, measured with observed OH sky lines.
The typical seeing size was $\sim 0\farcs9$ and the total integration time is 720 s (i.e., 120 s $\times$ six frames).

\begin{figure} % IR SPECTRUM -----------------------------------------------------------------------
\centering
\includegraphics[width=9.0cm]{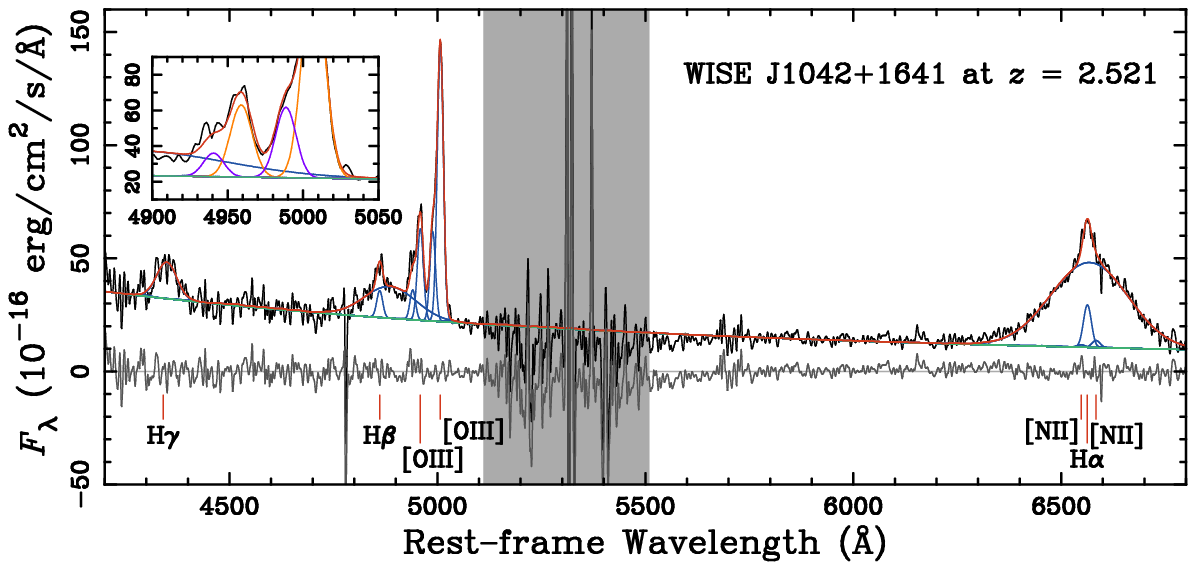}
\caption{Final reduced spectrum of WISE J1042$+$1641 observed with LIRIS shown as the black line. Red vertical lines indicate the central wavelengths of rest-frame optical emission lines, i.e., \hg, \hb, \oiii4959, \oiii5007, \nii6548, \ha,\ and \nii6583. The best-fit model for the observed spectrum is indicated as the red line, and the residual is shown as a dark gray line. The continuum (green line) and individual emission-line components (blue lines) are also shown. The gray shaded area was excluded in the fit. The small inset panel shows the same spectrum zoomed in to focus on the blueshifted outflow components in oxygen lines: blueshifted and unshifted lines are indicated as purple and orange lines, respectively.}
\label{f:irsp}
\end{figure} % -------------------------------------------------------------------------------------

Standard data reduction procedures were performed with the available IRAF tasks.
We used dome-flat frames to flat-field the object frames, and the first sky subtraction was executed with the A$-$B method.
The wavelength was calibratedwith arc (Ar and Xe) lamp frames.
We extracted the one-dimensional spectra with the {\tt apall} task using a 12-pixel ($3\farcs0$) aperture: moreover, we checked the two-dimensional spectrum to verify that no nearby sources within the extraction aperture contributed.
In this extraction, we fit the residuals of sky background and subtracted them.
The flux calibration and telluric absorption correction were carried out by using observed spectrum of a telluric standard star (HIP 30155).
We corrected the Galactic extinction with $E(B-V) = 0.025$ \citep{2011ApJ...737..103S}.
To correct observational uncertainties such as the slit loss and weather condition, we recalibrated fluxes by using $H$- and $K$-band photometric magnitudes obtained in the Two Micron All Sky Survey \citep[2MASS;][]{2006AJ....131.1163S}: the scaling factor is $\sim 3.1$.
As described in Sect.~3, our object is an anomalous gravitationally lensed object, showing four faint ($\sim$ 5--10\% fluxes of WISE J1042$+$1641, respectively) sources within $\sim$ 1\farcs6. Thus, they are contaminated in 2MASS low-resolution (2\farcs0) photometric magnitudes by $\sim$ 20--40\%: this does not affect our argument that the $M_{\rm BH}$/$M_\star$ ratio is significantly higher than the local relation.
After converting the spectrum into the rest frame, we also corrected for the intrinsic extinction adopting $E(B-V) = 0.71$, which was estimated in \citet{2018arXiv180705434G} from the AGN-origin continuum using a rest-frame UV-to-optical spectrum.
The reduced spectrum is shown in Fig.~\ref{f:irsp}.
We detected rest-frame optical emission lines, for instance, \hb, \oiii4959, \oiii5007, and \ha.

To extract the properties of the LIRIS spectrum, we fit the following spectral model.
We used the spectral wavelength range at $4200\AA < \lambda_{\rm rest} < 6800\AA$ for the fit, excluding a gap between $H$ and $K$ bands at $5110\AA < \lambda_{\rm rest} < 5510\AA$.
For the continuum emission, we adopted the power-law function and further included broad Fe emission components based on an empirical template \citep{2004A&A...417..515V}, although the iron contribution seems to be negligible for this object.
We adopted the Gaussian function to fit the line profiles.
The narrow components were fit simultaneously with the best-fit values of the velocity width and shift.
The flux ratios of the \nii6548,6583 and \oiii4959,5007 lines were fixed at the laboratory values of 2.96 and 3.03, respectively.
Because outflow components for the oxygen lines are present, we added another set of the \oiii4959,5007 lines with a fixed flux ratio.
We fit with a single Gaussian component for broad Balmer lines individually.
Figure~\ref{f:irsp} shows the best-fit model, and Table~\ref{t:prop} summarizes the fitting results.
Based on the spectral fitting, the redshift of WISE J1042$+$1641 is determined as $z = 2.5206\pm0.0001$: the uncertainty on the redshift reflects the fitting error.
We can estimate the central black hole mass by adopting a single-epoch method (see Sect.~\ref{disc}), since broad Balmer lines have been detected.
WISE J1042$+$1641 shows broad emission lines, even though it is a so-called dust-obscured galaxy.
This might be explained if the obscuring dust of DOGs is, as expected, clumpy rather than smooth.
If broad-line regions (BLRs) in DOGs suffer from such partial absorption, the intrinsic BLR and continuum luminosities increase by a factor of $1/(1-f)$, where $f$ is a covering factor.
If, for instance, $f = 0.8$, the difference is a factor of 5, corresponding to an $M_{\rm BH}$ difference by a factor of $\sim 5^{1/2}$.
This does not affect our argument.
We also detected blueshifted outflow components in the \oiii4959,5007 lines with a velocity offset of 1096.1 km/s.

\subsection{X-ray observation with Swift/XRT}

To obtain X-ray properties, we also observed WISE J1042$+$1641 using the Swift/X-Ray Telescope \citep[XRT;][]{2000SPIE.4140...64B} five times in 07--19 March 2018 with three-day intervals.
The net exposure was $\sim 9.2$ ks in total.
The data were reprocessed with the latest Swift calibration database (CALDB) as of March 22, 2018, through the script {\tt xrtpipeline} included in HEASoft version 6.21.
The source and background regions were defined as circles with a radius of 30\arcsec\ centered at the target position, and with a 280\arcsec\ radius in a nearby source-free area, respectively.
We merged all datasets in the five observations to produce the time-averaged spectrum.
We used {\tt swxpc0to12s6\_20130101v014.rmf} in CALDB as the response matrix file (RMF).
The ancillary response file was produced via {\tt xrtmkarf}.
For comparison, we also reduced the archival XRT data obtained in October 2011 and July 2012 in the same manner as the 2018 data, and produced a time-averaged (4.7 ks) spectrum.
The RMF {\tt swxpc0to12s6\_20110101v014.rmf} was employed for the spectrum in 2011 and 2012.

\begin{figure} % X-RAY SPECTRUM --------------------------------------------------------------------
\centering
\includegraphics[width=9.0cm]{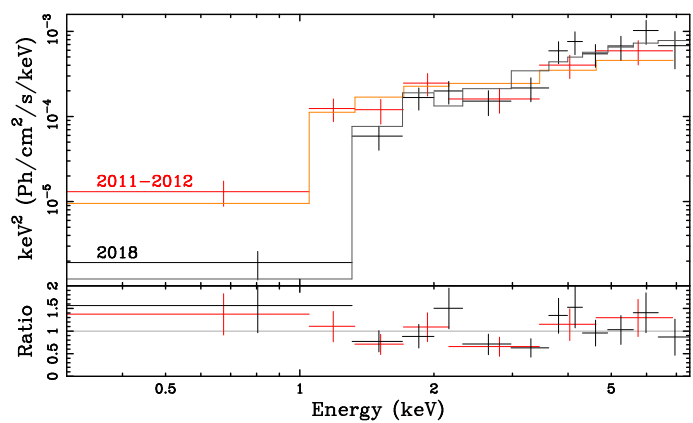}
\caption{Swift/XRT spectra in 2011--2012 (red) and in 2018 (black), with their best-fit absorbed {\tt pexmon} model shown as orange and dark gray lines, respectively ({\it top}) and the data-model ratios ({\it bottom}).
They are corrected for the effective area of the instrument and are given in units of $E F_E$ ($F_E$ is the energy flux at the energy $E$).}
\label{f:xrsp}
\end{figure}

Figure~\ref{f:xrsp} presents the time-averaged XRT spectra in 2011--2012 and 2018.
The source was found to change its spectral shape in these two periods.
To understand the origin of this variation, we fit these spectra with XSPEC version 12.9.1.
We first adopted an absorbed power-law model.
We modeled the Galactic absorption with {\tt phabs}, and the intrinsic absorption with {\tt zphabs}.
The Galactic column density, $N_\mathrm{H}^\mathrm{Gal}$, was set to be $2.2 \times 10^{20}$ cm$^{-2}$ (which was calculated with the FTOOLS {\tt nh} for the target position, based on the Galactic \ion{H}{i} map by \citealt{2005A&A...440..775K}), while the intrinsic one, $N_\mathrm{H}^\mathrm{int}$, was varied.
We obtain $N_\mathrm{H}^\mathrm{int}$ of $5.4^{+7.7}_{-5.3} \times 10^{22}$ cm$^{-2}$ and $21^{+7}_{-5} \times 10^{22}$ cm$^{-2}$, and photon indices ($\Gamma$) of $1.2\pm 0.5$ and $0.7^{+0.7}_{-0.6}$ from the 2011--2012 and 2018 data, respectively.
The resultant $\Gamma$ values are far lower than the canonical values in AGNs ($\sim$1.8), suggesting that the reflection component significantly contributes to the spectra.
Considering this result, we next replaced the power-law model with the {\tt pexmon} model, a cutoff power-law, and its reflection component from neutral matter including fluorescence lines \citep{2007MNRAS.382..194N}.
The redshift, cutoff energy, inclination angle, and $\Gamma$ were fixed at 2.521 (derived from our infrared spectrum), 100 keV, $60^\circ$, and 1.8, and the solid angle of the reflector normalized by $2\pi$ was varied within 0--2, a typical range observed in obscured AGNs \citep[e.g.,][]{2016ApJS..225...14K}.

The {\tt phabs*zphabs*pexmon} model is found to reproduce both spectra well (see Figure~\ref{f:xrsp}).
By modeling XRT spectra, we confirmed that the intrinsic luminosity (i.e., $\log L_{\rm 2-10 keV}$) remains little changed from 2011--2012 to 2018 (see Table~\ref{t:prop}).
In contrast, the intrinsic column density $N_\mathrm{H}^\mathrm{int}$ has significantly increased, from $9.4^{+5.8}_{-4.2} \times 10^{22}$ cm$^{-2}$ to $49^{+17}_{-13} \times 10^{22}$ cm$^{-2}$.
The relatively high $N_\mathrm{H}^\mathrm{int}$ values support the hypothesis that the source is an absorbed AGN.
The observed variation in $N_\mathrm{H}^\mathrm{int}$ is likely produced by an inner structure around the central black hole, such as the AGN torus.

 % -------------------------------------------------------------------------------------

\section{Discussion and conclusion}\label{disc}

Because broad Balmer lines are detected in the LIRIS spectrum, we estimated black hole masses with the single-epoch method as follows:
\begin{equation}
\log \left(\frac{M_{\rm BH}}{M_\odot}\right) = a + b \log \left(\frac{L}{10^{42} \ {\rm ergs \ s}^{-1}} \right) + c \log \left(\frac{{\rm FWHM}}{10^{3} \ {\rm km \ s}^{-1}} \right).
\end{equation}
For the case of the \ha\ line, we adopted a recipe using the \ha-line luminosity ($L_{\rm H\alpha}$) and its velocity width (FWHM$_{\rm H\alpha}$) with $a = 6.30$, $b = 0.55$, and $c = 2.06$ provided by \citet{2005ApJ...630..122G}. On the other hand, for the \hb\ line, we employed a recipe using the monochromatic continuum luminosity at 5100\AA\ ($L_{5100}$) and FWHM$_{\rm H\beta}$ with $a = 5.36$, $b = 0.64$, and $c = 2.00$ \citep{2005ApJ...630..122G}.
The estimated black hole mass is $\log (M_{\rm BH}/M_\odot) = 10.9$ for both cases (see Table~\ref{t:prop}).
Uncertainties of $M_{\rm BH}$ due to adopting different recipes \citep[e.g.,][]{2012ApJ...753..125S,2013ApJ...767..149B,2015ApJ...806..109J} are $\Delta \log M_{\rm BH} \sim 0.3$: when we adopt the recent recipes of \citet{2015ApJ...806..109J}, for example, black hole masses are derived as $\log (M_{\rm BH,H\alpha}/M_\odot) = 11.2$ and $\log (M_{\rm BH,H\beta}/M_\odot) = 10.8$.
We also calculated the averaged Eddington ratio of \ha- and \hb-based black hole masses, that is, $\log (L_{\rm bol,5100}/L_{\rm Edd}) = -0.323$, by estimating the AGN bolometric luminosity from the 5100\AA\ luminosity with a bolometric correction factor of 9.26 \citep{2006ApJS..166..470R}: when we use the 2$-$10 keV intrinsic X-ray luminosity with the conversion factor given by \citet{2004MNRAS.351..169M}, the bolometric luminosity and Eddington ratio are $\log L_{\rm bol,X} = 48.76$ and $\log (L_{\rm bol,X}/L_{\rm Edd}) = 0.34$, respectively, which is higher than the optical estimates (see Table~\ref{t:prop}).
As described in Sect.~2.2, we detected significant blueshifted outflow components of the \oiii4959,5007 lines.
This indicates that this DOG is in the blow-out phase at the end of the rapidly obscured black hole growth.
This is consistent with \citet{2017ApJ...850..140T}, who indicated that the \oiii4959,5007 velocity offset of most DOGs is larger than those of Seyfert 2 galaxies.

\begin{figure} % SED FITTING -----------------------------------------------------------------------
\centering
\includegraphics[width=9.0cm]{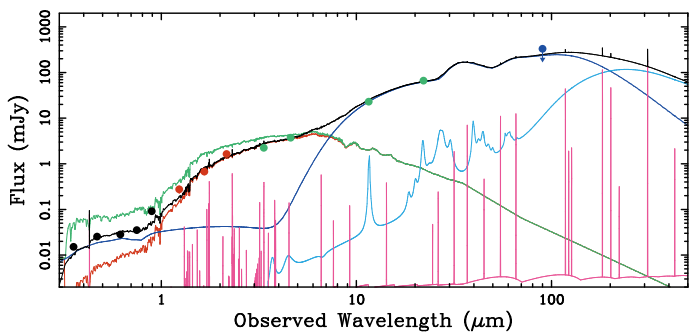}
\caption{SED fitting of WISE J1042$+$1641. Filled circles represent the observational data from the SDSS (black), 2MASS (red), and WISE (green). The 5$\sigma$ upper limit from AKARI is denoted as a blue arrow. The best-fit SED is denoted with the black solid line. Different components are also shown: unattenuated stellar emission (green line), attenuated stellar emission (red line), nebular emission (pink line), SF-heated dust emission (light blue line), and AGN emission (blue line).}
\label{f:xsed}
\end{figure} % -------------------------------------------------------------------------------------

In order to derive host-galaxy properties such as stellar mass and IR luminosity of WISE J1042$+$1641, we carried out the SED fitting with the code investigating galaxy emission \citep[CIGALE;][]{2005MNRAS.360.1413B,2009A&A...507.1793N}, which enables a SED modeling in a self-consistent manner by taking into account the energy equilibrium between the absorbed energy emitted in the UV/optical from SF/AGN and the re-emitted energy in IR from dust.
The SED fitting with CIGALE requires many parameters about the star formation history (SFH), single stellar population (SSP), attenuation law, AGN emission, and dust emission.
We applied a ``delayed'' SFH model that is defined as ${\rm SFR}\,(t) \propto t \times e^{-t/\tau}$ , where $t$ is the time and $\tau$ is the e-folding time of the old stellar population \citep[see, e.g.,][]{2016A&A...585A..43C}.
The age of oldest stars in the galaxy was parameterized with a range of 2.5--12.0 Gyr while the e-folding time was constrained in the range of 250--8000 Myr.
For the SSP and attenuation law, we adopted the stellar templates of \cite{2003MNRAS.344.1000B} with the \cite{2000ApJ...533..682C} dust extinction law assuming a \cite{2003PASP..115..763C} initial mass function (IMF).
We also added the standard default nebular emission model included in CIGALE.
We assumed solar metallicity ($Z = 0.02$) for the SSP model, and for the dust attenuation, the color excess of the stellar continuum, $E(B-V)$, was constrained within the range of 0.1--2.0.
We used the AGN model provided by \cite{2006MNRAS.366..767F} for the AGN emission \citep[see][for more detail]{2015A&A...576A..10C,2018ApJ...000L...0T}.
For the dust emission, we used the dust model of \cite{2007ApJ...657..810D}.
The fitting parameters were chosen based on our experiences of SED fitting for DOGs, some of which have a similar optical/IR color as WISE J1042$+$1641 \citep[e.g.,][]{2016ApJ...820...46T,2017ApJ...840...21T,2018ApJ...000L...0T}.
Therefore, our method with CIGALE is applicable to this object.
Under the above configuration with CIGALE, the possible fitting range of stellar mass in this SED fit is $8.72 < \log(M_\star/M_\odot) < 14.02$, which is a sufficiently wide range.
In this fitting, we considered SDSS ($u'$, $g'$, $r'$, $i'$, $z'$), 2MASS ($J$, $H$, $K_s$), WISE (3.4, 4.6, 12, 22 $\mu$m), and AKARI (90 $\mu$m) data.
We confirmed that WISE J1042$+$1641 is unresolved at all bands used in this work, which is reasonable because the angular resolutions of the SDSS, 2MASS, and WISE are higher than the size of a lensed system (see the next paragraph).
This means that the measured flux density at any band traces the total flux density of the lensed system.
The fitting results are shown in Figure~\ref{f:xsed}.

\begin{figure*} % M-SIGMA RELATION -----------------------------------------------------------------
\centering
\includegraphics[width=14.0cm]{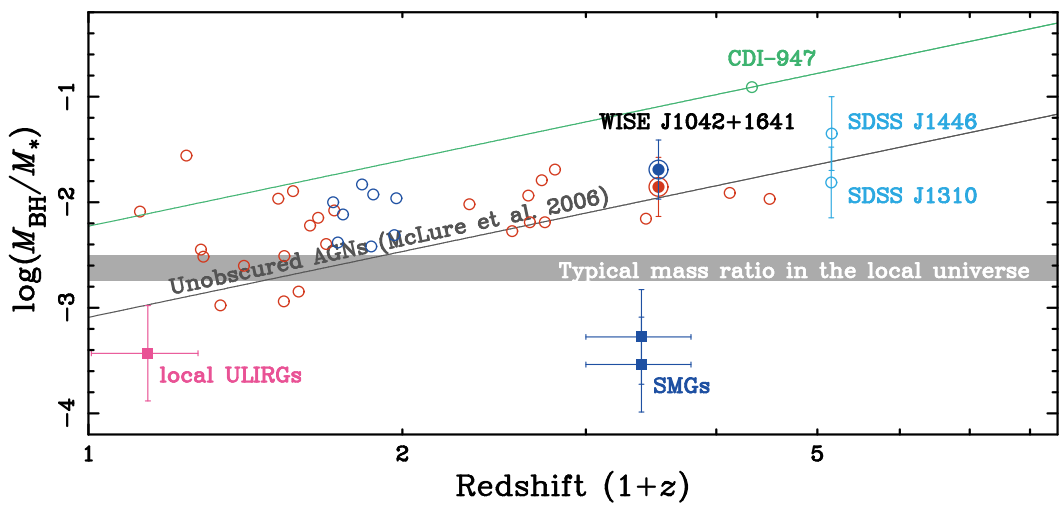}
\caption{Ratio of black hole to galaxy mass as a function of redshift. The $M_{\rm BH}/M_\star$ ratios of WISE J1042$+$1641 are shown as double circles and are color-coded according to the magnification factors of $\mu = 53$ (red) and 122 (blue). The black solid line indicates the $M_{\rm BH}/M_\star$ sequence of radio-loud AGNs in \citet{2006MNRAS.368.1395M}, followed by unobscured AGNs shown as red and blue open circles for lensed and non-lensed objects, respectively \citep{2006ApJ...649..616P}. The blue and pink filled squares denote X-ray obscured SMGs and local ULIRGs, respectively \citep{2008AJ....135.1968A}. SDSS luminous quasars at $z \sim 4$ (i.e., SDSS J1310$-$0055 and SDSS J1446$-$0101) in \citet{2012MNRAS.420.3621T} are denoted with light blue open circles. The green open circle denotes an X-ray selected unobscured AGN at the massive end, CDI-947 \citep{2015Sci...349..168T}, and the green line is a scaled sequence of \citet{2006MNRAS.368.1395M} to the source. The gray area is the typical mass ratio in the local Universe \citep{2003ApJ...589L..21M}.}
\label{f:msig}
\end{figure*} % ------------------------------------------------------------------------------------

We obtained the stellar mass ($M_\star$) and IR luminosity ($L_{\rm IR}$), listed in Table~\ref{t:prop}: the star formation rate was not determined due to the lack of constraints from the far-infrared data.
As shown in the table, an extremely high IR luminosity of $\log (L_{\rm IR}/L_\odot) = 14.57$, satisfying the criterion for extremely luminous IR galaxies (ELIRGs), and a stellar mass of $\log (M_\star/M_\odot) = 13.55$ were obtained.
In particular, this stellar mass seems to deviate from the known masses of $\log (M_\star/M_\odot) < 12$ \citep[e.g.,][]{2009ApJ...698..198G}.
Moreover, we checked its probability distribution function (PDF) by searching for a secondary peak, which we failed to find, however. This implies that the stellar mass is uniquely determined.
Very recently, based on high-resolution $J$- and $H$-band (F125W and F160W) imaging data obtained with the Wide Field Camera 3 (WFC3) on the {\it Hubble Space Telescope}, \citet{2018arXiv180705434G} reported that this object is an anomalous gravitationally lensed quasar, accompanied by four faint ($\sim$ 5--10\% respective fluxes of WISE J1042$+$1641) sources in the neighborhood (within $\sim 1.6$\arcsec), which looks like a quadruply lensed system.
They derived its lensing magnification factor of $\mu =$ 53--122: the magnification-corrected parameters of black hole mass, stellar mass, and various luminosities are listed in Table~1.
In this work, we discuss only the ratio of the black hole to galaxy mass, which is less strongly affected by the magnification correction, instead of absolute values of luminosities and masses.
We assumed the same magnification factor for AGN and stellar emission \citep[see also][]{2016MNRAS.458....2R,2006ApJ...649..616P}.
Although the compact AGN component could be more highly magnified than the host, this would not drastically change our considerations.

In Figure~\ref{f:msig} we plot the ratio of the black hole to galaxy mass of WISE J1042$+$1641.
We adopted the two magnification factors of $\mu = 53$ and 122 that were estimated in \citet{2018arXiv180705434G}.
We expect that the stellar mass at $z \sim 2$ and the bulge mass at $z = 0$ are comparable since WISE J1042$+$1641, the star-forming galaxy with $\log (M_{\rm BH}/M_\odot) \sim 10$ at $z \sim 2$, is likely the progenitor of local bulge-dominated galaxies \citep[e.g.,][]{2014A&A...567A.103B}.
As shown in the figure, we found that the ratio of the black hole to galaxy mass of WISE 1042$+$1641 (i.e., 0.0083--0.0120) is significantly higher than the ratio in the local Universe, which is $\log (M_{\rm BH}/M_\star) \sim$ 0.001--0.002 \citep[e.g.,][]{2003ApJ...589L..21M,2013ARA&A..51..511K}.
Unless the difference in magnification factors between AGN and host components (i.e., $\mu_{\rm AGN}/\mu_\star$) is larger than about eight, the $M_{\rm BH}/M_\star$ ratio lies above the local relation.
Our data points follow the sequence of unobscured AGNs \citep[e.g.,][]{2006MNRAS.368.1395M,2006ApJ...649..616P,2012MNRAS.420.3621T}, which are represented with a gray line, instead of SMGs \citep{2005ApJ...635..853B,2008AJ....135.1968A}, which are shown as squares.
This result suggests that WISE J1042$+$1641 is in a blow-out phase at the end of the buried rapid black hole growth, by considering its high Eddington ratios, \oiii4959,5007 outflow components, and the moderately high column density derived from X-ray spectra.
This is consistent with the recent result by \citet{2018ApJ...852...96W}, who suggested that a population of hyperluminous dusty galaxies, the so-called hot DOGs, represents a transitional high-accretion phase between obscured and unobscured quasars \citep[see also][]{2015ApJ...804...27A}.
Furthermore, the black hole mass of WISE J1042$+$1641 is quite high, even assuming the high magnification factor of $\mu = 122$: in this case, we obtain $\log M_{\rm BH,H\alpha} = 9.77$ with the recipe reported in \citet{2005ApJ...630..122G}.
Given the high black hole mass, this object should be the progenitor of a local massive galaxy that evolves to reach the massive end of the sequence of X-ray selected unobscured AGNs, like the source CID-947 studied by \citet{2015Sci...349..168T}, which is shown as a green circle and line in Figure~\ref{f:msig}.
However, to discuss the obscured evolution of  the ratios of black hole to galaxy masses in detail, additional data sets of dusty obscured populations are needed.

\begin{acknowledgements}
We would like to thank the anonymous referee for the useful comments and suggestions.
We are also grateful to Masamune Oguri, Yuichi Higuchi, Cristian E. Rusu, and Eilat Glikman for helpful comments and suggestions.
We thank the WHT staff, especially Ovidiu Vaduvescu and Raine Karjalainen, for invaluable support for the observation.
We also thank the Swift operation team for carrying out the X-ray observations.
The data analyses were in part carried out on the common-use data analysis computer system at the Astronomy Data Center, ADC, of the National Astronomical Observatory of Japan (NAOJ).
K. Matsuoka is supported by Japan Society for the Promotion of Science (JSPS) Overseas Research Fellowships.
This work is also supported by JSPS KAKENHI Grant Nos. 18J01050 (Y. Toba), 16H01101, 16H03958, 17H01114 (T. Nagao), 16K17672 (M. Shidatsu), 17K05384 (Y. Ueda), 15H02070, 16K05296 (Y. Terashima), 15K05030 (M. Imanishi).
Moreover, Y. Toba and W.-H. Wang acknowledge the support from the Ministry of Science and Technology of Taiwan (MOST 105-2112-M-001-029-MY3).
K. Iwasawa acknowledges support by the Spanish MINECO under grant AYA2016-76012-C3-1-P and MDM-2014-0369 of ICCUB (Unidad de Excelencia ``Mar\'ia de Maeztu'').
\end{acknowledgements}

\begin{table*} % ------------------------------------------------------------------------------------
\caption{General properties of WISE J1042$+$1641 measured from multiwavelength data set.}
\label{t:prop}
\centering
\begin{tabular}{llrrrrc}
\hline\hline
Parameter & Unit & Original Value && \multicolumn{2}{r}{Magnification-Corrected Values} & Source$^b$\\
          &      & $\mu = 1$      && $\mu = 53$ & $\mu = 122$                           &\\
\hline
$z_{\rm spec}$            & --                                & 2.5206$\pm$0.0001    && --                   & --                   & 1\\
$F_{\rm H\alpha}$         & 10$^{-16}$ erg s$^{-1}$ cm$^{-2}$ & 8741.96$\pm$98.55    && 164.94$\pm$1.86      & 71.66$\pm$0.81       & 1\\
FWHM$_{\rm H\alpha}$      & km s$^{-1}$                       & 9995.18$\pm$122.66   && --                   & --                   & 1\\
$\log L_{\rm H\alpha}$    & erg s$^{-1}$                      & 46.65$\pm$0.01       && 44.93$\pm$0.01       & 44.56$\pm$0.01       & 1\\
$F_{\rm H\beta}$          & 10$^{-16}$ erg s$^{-1}$ cm$^{-2}$ & 2177.61$\pm$93.11    && 41.09$\pm$1.76       & 17.85$\pm$0.76       & 1\\
FWHM$_{\rm H\beta}$       & km s$^{-1}$                       & 8951.40$\pm$436.34   && --                   & --                   & 1\\
$\log L_{\rm H\beta}$     & erg s$^{-1}$                      & 46.04$\pm$0.04       && 44.32$\pm$0.04       & 43.95$\pm$0.04       & 1\\
$\log L_{\rm 5100}$       & erg s$^{-1}$                      & 47.73$\pm$0.01       && 46.01$\pm$0.01       & 45.65$\pm$0.01       & 1\\
$\log L_{\rm bol,5100}$   & erg s$^{-1}$                      & 48.70$\pm$0.01       && 46.98$\pm$0.01       & 46.61$\pm$0.01       & 1\\
$\log L_{\rm 2-10keV}$    & erg s$^{-1}$                      & 46.5$\pm 0.1$        && 44.8$\pm 0.1$        & 44.4$\pm 0.1$        & 2\\
$\log L_{\rm 2-10keV}$    & erg s$^{-1}$                      & 46.7$_{-0.1}^{+0.2}$ && 45.0$_{-0.1}^{+0.2}$ & 44.6$_{-0.1}^{+0.2}$ & 3\\
$R$$^a$                   & --                                & $2.0_{-1.2}^{+0.0}$  && --                   & --                   & 2\\
$R$$^a$                   & --                                & $2.0_{-1.4}^{+0.0}$  && --                   & --                   & 3\\
$N_{\rm H}^{\rm int}$     & 10$^{22}$ cm$^{-2}$               & 9.4$_{-4.2}^{+5.8}$  && --                   & --                   & 2\\
$N_{\rm H}^{\rm int}$     & 10$^{22}$ cm$^{-2}$               & 49$_{-13}^{+17}$     && --                   & --                   & 3\\
$\log L_{\rm bol,X}$      & erg s$^{-1}$                      & 49.36$\pm$0.23       && 47.64$\pm$0.23       & 47.27$\pm$0.23       & 3\\
$\log M_{\rm BH,H\alpha}$ & $M_\odot$                         & 10.92$\pm$0.03       &&  9.97$\pm$0.03       &  9.77$\pm$0.03       & 1\\
$\log M_{\rm BH,H\beta}$  & $M_\odot$                         & 10.93$\pm$0.10       &&  9.83$\pm$0.10       &  9.60$\pm$0.10       & 1\\
$\log L/L_{\rm Edd,5100}$ & --                                & $-$0.32$\pm$0.03     && $-$1.09$\pm$0.03     & $-$1.25$\pm$0.03     & 1\\
$\log L/L_{\rm Edd,X}$    & --                                & 0.34$\pm$0.23        && 1.29$\pm$0.23        & 1.49$\pm$0.23        & 1 \& 3\\
$\log L_{\rm IR}$         & $L_\odot$                         & 14.57$\pm$0.11       && 12.85$\pm$0.11       & 12.48$\pm$0.11       & 4\\
$\log M_\star$            & $M_\odot$                         & 13.55$\pm$0.02       && 11.82$\pm$0.02       & 11.46$\pm$0.02       & 4\\
\hline
\end{tabular}
\tablefoot{$^a$ the solid angle of the reflector, normalized by $2\pi$: the upper limit was pegged, $^b$ source: $1 =$ LIRIS spectrum, $2 =$ 2011--2012 XRT spectrum, $3 =$ 2018 XRT spectrum, and $4 =$ SED fit.}
\end{table*} % --------------------------------------------------------------------------------------

\end{document}